\documentclass[%
 reprint,
 amsmath,amssymb,
 aps,
]{revtex4-1}

\usepackage{graphicx}
\usepackage{dcolumn}
\usepackage{bm}
\usepackage[T1]{fontenc}

\begin{document}


\title{A quantum analogue to the deflection function}

\author{P. G. Jambrina}
 \email{pablo.gjambrina@uam.es}
\affiliation{%
 Departamento de Qu\'{\i}mica F\'{\i}sica Aplicada,
Universidad Autonoma de Madrid, 28049, Madrid, Spain\
}%
\author{M. Menéndez}
\author{F. J. Aoiz}
 \email{aoiz@ucm.es}
 \affiliation{
 Departamento de Qu\'{\i}mica
F\'{\i}sica I, Facultad de Ciencias Qu\'{\i}micas, Universidad Complutense de
Madrid , 28040 Madrid, Spain
}%

\date{\today}

\begin{abstract}
The classical deflection function is a valuable computational tool to
investigate reaction mechanisms. It provides, at a glance, detailed information
about how the reaction is affected by changes in reactant properties (impact
parameter) and products properties (scattering angle), and, more importantly,
it also shows how they are correlated. It is also useful to predict the
presence of quantum phenomena such as interferences. However, rigorously
speaking, there is not a quantum analogue as the differential cross section
depends on the coherences between the different values of $J$ caused by the
cross terms in the expansion of partial waves. Therefore, the classical
deflection function has a limited use whenever quantum effects become
important. In this article, we present a method to calculate a quantum
deflection function that can shed light onto reaction mechanism using just
quantum mechanical results. Our results show that there is a very good
agreement between the quantum and classical deflection function as long as
quantum effects are not all relevant. When this is not the case, it will be
also shown that the quantum deflection function is most useful to observe the
extent of quantum effects such as interferences. The present
results are compared with other proposed quantum deflection
functions, and the advantages and disadvantages of the different formulations will be discussed.
\end{abstract}

\maketitle


\section{Introduction}

The main goal of reaction dynamics is to obtain the various microscopical
properties as excitation functions or rotational distributions and from them,
macroscopical properties such as thermal rate coefficients. Overall, the
process is equivalent to disentangling how microscopical properties govern the
macroscopic outcome. Accordingly, it is not enough to reproduce and predict
experimental measurements, but it is also important to unveil the detailed
reaction mechanisms.

Impact parameter $b$ (or orbital angular momentum $\bm \ell$) and scattering angle
$\theta$ are two of the main variables that are studied to discern reaction mechanisms.
The former is related to the reactants asymptote, and is one of the key players in
determining the outcome of a collision \cite{Levinebook,Tutorialsbook} as it determines
the parts of the potential energy surface (PES) that will be explored during the
collision (head-on {\em vs.} glacing collisions). The scattering angle, in turn, is
defined at the product asymptote and provides information about nuclei scrambling during
the collision; besides, it is amenable of experimental measurement using cross molecular
beams with mass spectrometric universal detection or, more recently, velocity-mapped ion
imaging \cite{Parker:RSI1998,GLRWS:JCP05,Liu:RSI2003,Ashfold:PCCP2006} or single beam
coexpansion such as photoloc \cite{SOSXZ:CPL93} among other techniques. Moreover, it is
relatively straightforward to extract the reaction (or inelastic) probability as a
function of $J$ (opacity function or $P_r(J)$), and the differential cross section (DCS)
as a function of the scattering angle $\theta$. Hence, it is not surprising that $P_r(J)$
and DCS are two of the most important observables in reaction dynamics.

The deflection function, that is, is the joint dependence of the reaction probability as
a function of the scattering angle and the impact parameter, contains all the information
provided by the $P_r(J)$ and the DCS and, above all, how $J$ and $\theta$ correlate
throughout the collision. The classical deflection function has been widely used to
explain elastic and inelastic scattering, in particular to understand those features
related with glory and rainbow scattering
\cite{Ford-Wheeler:AP59,Bernstein:1966,Levinebook}. For reactive scattering a strong
correlation between $J$ and $\theta$ is expected for reactions following a direct
mechanism, whereas none or very weak correlation between these variables can be
anticipated if the reaction takes place through a long-lived collision complex.
Furthermore, discontinuities and different trends in the deflection function can be used
to characterized different reaction mechanism even for apparently simple
reactions\cite{GMW:JCP08,GMWA:JCP08}.

The classical deflection function has been also used to predict interferences causing
oscillations in the DCS. Given the wave nature of quantum mechanics (QM), it is expected
that when one particle may follow two different pathways giving rise to the same outcome,
they will interference. In the double-slit Young experiment \cite{Feynmanbook}
interferences arise when electrons going through two different slits could hit the
detector. In reaction dynamics we do not need slits and the system itself acts as an
interferometer whenever two different $J$ could scatter at the same angles
\cite{JHASJZ:NC15,JAASZ:CS16,SGZJMA:JCP16}. This analogy also explains why the deflection
function cannot be calculated using pure quantum mechanical grounds in the same way
as it is done in classical calculations. In quantum mechanics, the angular distribution
depends on the coherences between different $J$-partial waves, therefore something
apparently as simple as obtaining a rigorous joint probability distribution as a
function of $J$ and $\theta$ cannot be computed. That would be similar to disentangle
which parts of the signal comes from electrons going through one or other slit in the
Young double-slit experiment.

It is not surprising that many efforts have been done to overcome this limitation.
Especially interesting is the so-called \textit{Quantum Deflection Function} (CQDF)
devised by Connor and coworkers \cite{C:PCCP04,SC:PCCP11,XC:JPCA09} in the context of the
glory analysis of forward scattering. The CQDF, defined as the derivative of the argument
of the scattering matrix element with respect to $J$, has probed to be a valuable tool to
predict the presence of rainbows and to identify the rainbow angular momentum variable.
Besides, it could be used to predict interferences between nearside and farside
scattering. However, CQDF provides a single value of $\theta$ (actually, of the
deflection angle, $\Theta$, whose absolute value is the measurable scattering angle
$\theta$) for one or few $J$s so it cannot be considered as a joint probability of $J$
and $\theta$. Moreover, the CQDF does not consider that a single $J$ can correlate with a
continuous series of different $\theta$ which impairs its use to predict the presence of
different mechanisms.

Throughout this article, we will try to circumvent this limitation, and we will propose a
new quantum analogue to the classical deflection function, $Q_r(\theta,J)$ or QM-DF,
which may be useful for the interpretation of quantum scattering results. This new
function is a sort of joint distribution of $J$ and $\theta$ that includes all
coherences between different partial waves, and whose summation over all partial waves
recovers the exact angular distribution. The article is organized as follows: in Section
II we will revise the classical deflection function as the joint distribution of $\theta$
and $J$, followed by the definition of a intuitively simple QM-DF, $Q_r(\theta, J)$,
starting from the definition of the scattering amplitude, that will be compared with the
QCT deflection function. In Section III we will assay the validity and usefulness of the
proposed QM-DF for three different systems and situations. First of all, we will study
the inelastic collisions of Cl + H$_2$, where the QCT deflection function succeeded
in explaining the quantum results. Next, we will study the reactive D$^+$ + H$_2$ system,
prototype of barrierless reactions where we expect no correlation between $J$ and
$\theta$. Finally, we will apply the QM-DF to reactive scattering between H and D$_2$ at
high collision energies where quantum interferences govern the angular distributions for
certain combinations of final and initial states. For all these systems, QM calculations
have been carried out using the close-coupling hypespherical method of Skouteris
\textit{et al.} \cite{SCM:CPC00}, while QCT calculations have been performed using the
procedure described in Refs.~\citenum{ABH:JCSFT98} and \citenum{AHR:JCP92}.

\section{Theory}

\subsection{Classical Deflection Function}

The basis of the QCT method consists in calculating an ensemble of trajectories following
a judicious sampling of initial conditions to cover as much as possible the phase space
relevant for the process to be studied but complying with the state quantization of
the reactants. The initial and final atom positions and linear momenta are then used to
determine those initial and final properties (such as angular momenta, scattering angle,
final states, etc.) necessary to characterize each individual trajectory. Finally, all is
needed is to carry out the average value of any conceivable property over the ensemble of
trajectories. For example, the total reaction probability for a given value of the total
angular momentum quantum number, $J$, discretely sampled can be obtained as:
\begin{equation}\label{pj1}
  P_r(J)=\frac{N_r(J)}{N_{\rm tot}(J)}
\end{equation}
where $N_r(J)$, and $N_{\rm tot}(J)$ are the number of reactive (or inelastic if that
were the case) and total trajectories, respectively, for a given $J$. Recall that the
total angular momentum $\bm J=\bm \ell+\bm j$, where $\bm j$ is the rotational angular
momentum and $\bm \ell$ is the (relative) orbital angular momentum. We can define the
corresponding quantum numbers, $J$, $\ell$ and $j$, such that ${|\bm J|=[J(J+1)]^{1/2}
\hbar}$ and similarly for $|\bm \ell|$ and $|\bm j|$. These quantum numbers can be
sampled continuously (real values) or discretely (integer values).

Equation~\eqref{pj1} is valid if the sampling in $J$ is discretely and
uniformly sampled and, similarly, for the orbital angular momentum in the
{$|J-j| \le \ell \le J+j$} interval (for details see ref.
\citenum{ASMG:JCP05}). In addition, not all reactive trajectories need to have
the same weight. Sometimes it is necessary to attribute different weights to
each trajectory as is done in the Gaussian binning
procedure\cite{BR:JCP97,BAHL:JCP03,BR:CPL04} to make the assignment of final
rovibrational states `more quantal',  or simply because a biased sampling is
used. In those cases, $N_r(J)$ in Eq.~\eqref{pj1} is replaced by $S_w$, the sum
of the weights of reactive (or inelastic) trajectories into a given final
manifold of states. If one wishes to calculate a property that depends on more
than one variable, for example of $J$ and $\ell$, the scheme is the same except
that now a joint probability has to be considered (say, the number of reactive
trajectories with values of $J$ and $\ell$, $N_r(J,\ell)$).\cite{ASMG:JCP05}
The aforementioned procedure is suitable for discrete variables, while for
continuous variables it is a common practice to use histograms or, more
elegantly, to fit the distributions to series of orthogonal polynomials.
\cite{AHS:JCP92,ABH:JCSFT98,ASMG:JCP05} Obviously, integration (or summation)
over one of the variables of a given joint probability distribution, leads to
the probability distribution of the other variable. Moreover, if we split the
original ensemble of trajectories in a series of sub-ensembles and calculate
the respective joint probability distribution, it turns out that the global
probability distribution can be easily  recovered from the joint probabilities
distributions for all the sub-ensembles; that is to say, the probability
distributions are always additive. As we will see, this is not the case in QM
scattering due to the interferences.

To illustrate the calculation of the classical deflection function, let us
assume that the orbital angular momentum is sampled continuously in the ${\ell
\in [0, \ell_{\max}]}$ with a weight $2\ell+1$, that is, the orbital angular
momentum for the $i$-th trajectory is sampled as $\ell_i(\ell_i+1)=\xi_i
[\ell_{\max}(\ell_{\max}+1)]$, where $\xi$ is a random number in $[0,1]$ (this
is the same as sampling the impact parameter as $b =\xi^{1/2} b_{\max}$).

We can conveniently define a $J$-partial cross section, $\sigma_r(J)$:
\begin{eqnarray}
\sigma_r(J)=\frac{\pi}{k^2} \,(2J+1) \frac{2\min(J,j)+1}{2j+1} P_r(J)\,,
\end{eqnarray}
which is nothing but a probability density function normalized such that the integral or
the sum of $\sigma_r(J)$ over $J$ is the integral cross section, $\sigma_r$, either total
or into a given final state.\cite{ASMG:JCP05} For discrete values of $J$, $\sigma_r(J)$
is usually denoted in the literature as $\sigma_r^J$.

The Monte Carlo normalized probability density function can be written as
\begin{eqnarray} \label{GJ}
\sigma_r(J)=\frac{\sigma_r}{S_w} \, \sum_{i=1}^{N_r} w_i \delta(J-J_i)\,,
\end{eqnarray}
where $w_i$ and $J_i$ are the weight and $J$ value of the $i$-th trajectory. $S_w$ is the
sum of the weights of all the relevant reactive trajectories, $S_w=\sum w_i$. In the
simplest case, $w_i$ would be a Boolean function whose value is one only for the specific
reactive trajectories and zero otherwise, such that $S_w=N_r$, the number of the
considered reactive trajectories. As a convenient approximation, the Dirac delta
functions can be replaced with a normalized Gaussian function
\begin{eqnarray}
G(J-J_i)=\frac{1}{s \sqrt{\pi}} \exp\left[- \frac{(J-J_i)^2}{s^2} \right] \, ,
\end{eqnarray}
where the width, $s=\Delta_{\rm FWHM}/\ln 2$, is conveniently chosen depending
on the average spacing of the successive values of $J_i$ and the statistical
uncertainty.

If the sampling in $J$ (and in $\ell$) is made continuous, the $J$-partial cross section
can be expressed as an expansion in Legendre polynomials, $P_n(x)$:
\begin{eqnarray} \label{LegJ}
\sigma_r(J)=\sigma_r \,\,\frac{2(2J+1)}{J_{\max}(J_{\max}+1)} \, \sum_{n} b_n
P_n[x(J)] \,,
\end{eqnarray}
where $x$ is a reduced variable, $x \in [-1,1]$, given by
\begin{eqnarray}
x=\frac{J(J+1)}{J_{\max}(J_{\max}+1)} -1 \,, \label{x}
\end{eqnarray}
where $J_{\max}$ is the maximum value of the total angular momentum used in the
calculation to ensure the convergence. The coefficients, $b_n$, are given in terms of the
Legendre moments as
\begin{eqnarray}
b_n= \frac{2n+1}{2} \, S_w^{-1}\,\sum_{i=1}^{N_r} w_i \, P_n(x_i)\,,
\end{eqnarray}
where $x_i$ is the value of $x$, given by Eq.~\eqref{x}, of the {\em i}-th
trajectory, and $P_n(x)$ is the n-th order Legendre polynomial.

Similarly, the  DCS can be expressed as an expansion in Legendre
polynomials:
\begin{equation}\label{dcsqct1}
\sigma_r(\theta) \equiv \frac{d\sigma(\theta)}{d\omega}= \frac{\sigma_r}{2 \pi}
\,\sum_{m=0} a_m P_m (\cos\theta),
\end{equation}
where $\sigma_r$ is the integral cross section, and $a_n$ are the expansion coefficients
whose values are given by:
\begin{equation} \label{dcsqct2}
a_m = \frac{2m+1}{2} \langle P_m(\cos\theta) \rangle = \frac{2m+1}{2}\,
S_w^{-1}\, \sum_{i=1}^{N_r} w_i P_n(\cos\theta_i)\,,
\end{equation}
where $\langle P_m(\cos\theta) \rangle$ is the weighted average value of
$P_m(\cos\theta)$ over the ensemble of the relevant trajectories.

The classical deflection function, that is, the joint probability distribution of $J$ and
$\theta$, normalized to the integral cross section, can now be expressed as a
double expansion in Legendre polynomials
\begin{eqnarray}\label{Legdef}
\sigma_r(\theta, J) &=& \frac{\sigma_r}{2 \pi} \,
\frac{2(2J+1)}{[J_{\max}(J_{\max}+1)]} \sin\theta \, \cdot \\ \nonumber & & \sum_{m=0} \, \sum_{n=0}
\alpha_{mn} P_m (\cos\theta) P_n[x(J)]
\end{eqnarray}
where the coefficients  $\alpha_{mn}$ are given by:
\begin{eqnarray} \label{clasdeflfun2}
&\alpha_{mn}&= \frac{(2m+1)(2n+1)}{4}  \langle P_m(\cos\theta)  P_n[x(J)]
\rangle=  \\ \nonumber && \frac{(2m+1)(2n+1)}{4} S_w^{-1} \sum_{i=1}^{N_r} w_i
P_m(\cos\theta_i) P_n[x_i(J_i)]
\end{eqnarray}

The Monte Carlo expression of the deflection function can be expressed as a sum
of Gaussian functions given by
\begin{eqnarray}\label{Gdeflec}
&\sigma_r(\theta, J)&= \frac{\sigma_r}{2\pi} \,S_w^{-1}~\sum_{i=1}^{N_r} w_i\,
\delta(J-J_i) \delta(\theta-
\theta_i) \approx \nonumber \\
&& \frac{\sigma_r}{2\pi} \, S_w^{-1} ~\sum_{i=1}^{N_r} w_i \, G(J-J_i)
G(\theta- \theta_i) \label{Drg}
\end{eqnarray}
where $J_i$ and $\theta_i$ represent the values of $J$ and $\theta$ for the
$i$-th trajectory. $G(J-J_i)$ and $G(\theta- \theta_i)$ denote normalized
Gaussian functions with width parameters $s_{_{J}}$ and $s_{\theta}$, centred
in $J_i$ and $\theta_i$, respectively.

Integration of Eq.~\eqref{Legdef} or Eq.~\eqref{Drg} over $\theta$ and the azimuthal
angle renders the $J$-partial cross section of Eq.~\eqref{LegJ} and Eq.~\eqref{GJ}.
Alternatively, integration over $J$ in those equations gives the $\sigma_r(\theta) \,
\sin \theta$.


\subsection{QM analogue to the Deflection Function}

Due to its classical nature, there is no restriction in QCT calculations to obtain any
correlation between two or more properties. After all, each trajectory is characterized
by specific values of any initial or final property. However, this is not the case for QM
scattering calculations, which makes the analysis based on pure QM calculations not so
trivial. From the QM scattering calculations we only obtain as an outcome the scattering
matrix (S-matrix) that relates the initial states of the reactants and the final states
of the products. In the unsymmetrized representation, the S-matrix has one element per
energy, chemical rearrangement $\alpha$, $J$, and initial and product states. For the
particular case of closed shell diatomic molecules in the helicity representation
(body-fixed frame), and a given value of $J$, these are characterized by three
quantum numbers for each arrangement: $v$, $j$, ($v'$ and $j'$) that define the
vibrational and rotational states respectively, and the helicity $\Omega$ ($\Omega'$),
the projection of $\bm j$ ($\bm j'$) (or $\bm J$) onto the approach (or recoil)
direction. It means that to obtain a dynamical observable from a QM calculation, we need
a recipe to extract its value from the elements of the S-matrix.

Some observables can be readily extracted from the S-matrix. This is the case of the
$P_r(J;E)$ that, for a given initial state and total energy, can be calculated as follows:
\begin{equation}\label{pjqm}
P_r(J;E) = \frac{1}{2\min(J,j)+1}\,\, \sum_{\Omega}\sum_{\Omega'}   \left|
S^{J \alpha}_{v' j' \Omega', v j \Omega} (E)\right|^2
\end{equation}
where the sum runs over the desired products states (or, if referred to state-to-state,
without summing over $v'$ and $j'$). Hereinafter,  subscripts for the $v$, $j$, $v'$,
$j'$, energy, and the chemical arrangement will be omitted for clarity. The integral
cross section can be written in terms of the reaction probabilities as
\begin{eqnarray}\label{sigJQM}
\sigma_r(E) &=& \frac{\pi}{k_{vj}^2} \, \sum_{v',j'} \, \sum_{J=0}^{J_{\max}}
(2J+1)\,\frac{2\min(J,j)+1}{j+1}\,\,  P_r(J;E)=\nonumber
\\&&\sum_{J=0}^{J_{\max}} (2J+1)\, \sigma_r^J(E)
\end{eqnarray}
where $k^2_{vj}=2 \mu (E-E_{vj})/\hbar^2$,  is the initial relative wavenumber vector,
and $\mu$ is the atom-diatom reduced mass. $J_{\max}$ is the maximum value of $J$
necessary for convergence. $\sigma_r^J(E)$ is the $j$-partial cross section already
mentioned in the previous subsection.

To extract vector properties such as the DCS from the S-matrix is not so
straightforward. First, because we need to include the angular dependence;
second, because they involve coherences between different elements of the
S-matrix. It is convenient to express the DCS in terms of the scattering
amplitudes, which are defined as:
\begin{equation}\label{scattampl}
f_{\Omega' \Omega} (\theta) = \frac{1}{2 \imath k_{vj}} \sum_{J=0}^{J_{\rm
max}} (2J+1) d^J_{\Omega' \Omega} (\theta) S^{J }_{ \Omega', \Omega}
\end{equation}
where $d^J_{\Omega' \Omega} (\theta)$ is the Wigner d-matrix. The DCS can now be written
using the scattering amplitudes as:
\begin{equation}\label{dcsqm1}
\sigma_r(\theta)\equiv \frac{d\sigma_r(\theta)}{d\omega}= \frac{1}{2 j + 1}
\sum_{\Omega' \Omega} f^*_{\Omega' \Omega} (\theta) f_{\Omega' \Omega} (\theta)
\end{equation}
From Eqs.~\eqref{scattampl} and Eq.~\eqref{dcsqm1} it is clear that the DCSs
for state-to-state processes are additive, even when they are resolved in
$\Omega'$, and $\Omega$. However, the squaring of the sum over $J$ in
Eq.~\ref{scattampl} makes the DCS no longer additive in $J$. This property is a
reflection of the wave nature of quantum mechanics, so that two ``paths''
(impact parameters or $J$) leading to scattering at the same angles interfere.
Hence, in principle, it is not possible to separate the contribution of two
mechanisms (or paths) in a overall DCS. It is worth noticing that usually the
interference are only important between nearby values of $J$
\cite{PHJAAA:PCCP12} so, for certain cases, it is possible to extract the
contributions from one or many mechanisms from the DCS.

To calculate a QM deflection function we would need to extract the contribution
of each $J$ to the total DCS. Furthermore, to be reliable and to provide a
valuable insight into the collision mechanism, the QM deflection function
should be additive, so that the sum over $J$ should be enough to recover the
overall DCS. One could, in principle, compute it by neglecting all coherences
between different $J$s. This would be equivalent of using the random phase
approximation that lies in the core of the statistical model
\cite{RGM:JCP03,RHM:CPL01}, giving rise to forward-backward symmetric DCSs. For
non-statistical (direct) reactions, a symmetric DCS is in clear disagreement
with the experimental results, and hence neglecting coherences can be
considered as a very unappropriate approximation to obtain a QM deflection
function. To devise a QM analogue to the deflection function we will start by
defining a $J$-partial dependent scattering amplitude as:
\begin{equation}\label{scattj}
f^J_{\Omega' \Omega} (\theta) = \frac{1}{2i  k_{vj}}   \, (2 J +1) d^J_{\Omega'
\Omega}(\theta) S^J_{\Omega' \Omega}
\end{equation}
where $|\Omega|,~|\Omega'| \le J$. The (total) scattering amplitude can now be
written as
\begin{equation}
f_{\Omega' \Omega} (\theta) =  \sum_{J=0}^{J_{\rm max}} f^J_{\Omega' \Omega}
(\theta)
\end{equation}
The DCS can be expressed as a function of the $J$-partial scattering
amplitudes:
\begin{equation}\label{dcsscattj1}
\sigma_r(\theta)=  \frac{1}{2 j + 1} \sum_{\Omega' \Omega}~
\sum_{J_1=0}^{J_{\max}} \sum_{J_2=0}^{J_{\max}} f^{J_1}_{\Omega' \Omega}
(\theta)  f^{J_2 *}_{\Omega' \Omega} (\theta) \,,
\end{equation}
which is the same as Eq.~\eqref{dcsqm1}. Without any approximation,
Eq.~\ref{dcsscattj1} can be rearranged to
\begin{eqnarray}\label{dcsscattj2}
\sigma_r(\theta) &=& \frac{1}{(2 j + 1)} \sum_{\Omega' \Omega} ~ \sum_{J=0}^{J_{\rm
max}}  {\ } \sum_{J_1=0}^{J_{\max}} \sum_{J_2=0}^{J_{\max}} \, \frac{\left(
\delta_{J_1,J} + \delta_{J_2,J}\right)}{2} \cdot \nonumber \\  & & f^{J_1}_{\Omega' \Omega}
(\theta) f^{J_2 *}_{\Omega' \Omega} (\theta).
\end{eqnarray}
Eqs.~\eqref{dcsscattj1} and \eqref{dcsscattj2} only differ in the presence of
an additional sum over $J$ in \eqref{dcsscattj2} that is compensated with the
term $(\delta_{J_1,J} + \delta_{J_2,J})/2$, that guarantees that both equations
include the same number of cross products and hence that they are equivalent.
The advantage of Eq.~\eqref{dcsscattj2} is the presence of a separate summation
over $J$ that allows us to define a function that depends on a single $J$ and
$\theta$; that is, a {\em quantum analogue} to the classical deflection
function (QM-DF) that we will denote as $Q_r (\theta,J)$,
\begin{eqnarray}\label{QJ}
Q_r (\theta, J) &=& \frac{\sin \theta}{2 j + 1} \, \sum_{\Omega' \Omega} ~
\sum_{J_1=0}^{J_{\max}} \sum_{J_2=0}^{J_{\max}} \, \frac{\left( \delta_{J_1,J}
+ \delta_{J_2,J}\right)}{2}  \cdot \nonumber \\ & & \, \, f^{J_1}_{\Omega' \Omega} (\theta)  f^{J_2
*}_{\Omega' \Omega} (\theta)\, .
\end{eqnarray}
To help the interpretation of the quantum deflection function defined in this
work, Eq.~\eqref{QJ} can be recast as
\begin{eqnarray}\label{QJ2}
Q_r (\theta, J) &=& \frac{\sin \theta}{2 j + 1} \, \sum_{\Omega' \Omega} ~
|f^{J}_{\Omega' \Omega}|^2 \cdot \nonumber \\ &+&  \frac{1}{2}\,   \sum_{\Omega' \Omega} \sum_{\substack{J_1=0 \\ J_1 \ne
J}}^{J{\max}} \, \left[f^{J}_{\Omega' \Omega} (\theta)  f^{J_1 *}_{\Omega'
\Omega} (\theta) +~ \text{c.c.} \right]   \, .
\end{eqnarray}
where c.c. stands for the respective conjugate complex. Equation~\eqref{QJ2}
contains the square of the $J$-dependent scattering amplitude, $|f^J_{\Omega'
\Omega} (\theta)|^2$, plus a halved summation of $J_{\max}$ terms over all the
total angular momenta $J_1\ne J$, which are the coherent terms. The other half
of the summation will appear in previous or subsequent values of $J$. In the
absence of coherences, that is, in the random phase approximation limit, the
only surviving term would be that depending of $J$ only. The remaining terms
account for the possible interferences that most of the time can be expected to
be only important between partial waves in a restricted range of $J$ in
$[J-\Delta J, J+\Delta J]$.\cite{JHASJZ:NC15,JAASZ:CS16} However, as it will be
shown below, interferences can also take place between partial waves that cover
the full range of angular momentum leading to scattering.

The QM-DF shares some important properties in common with the
classical ones. As in the classical case, summing Eq.~\eqref{QJ} over $J$ leads
to the DCS given by Eq.~\eqref{dcsscattj2} multiplied by $\sin \theta$,
$\sigma_r(\theta) \sin \theta$. Similarly, by integration over the scattering
angle and the azimuthal angle,
\begin{equation}\label{intQJ}
\int_{-1}^{1} {\rm d}\theta ~~Q_r (\theta, J) = \frac{\pi}{k^2_{v,j}}\, \,
\frac{2J+1}{2j+1} \sum_{\Omega' \Omega} ~ |S^J_{\Omega' \Omega}|^2 =
\sigma^J(E)\,,
\end{equation}
gives the $J$-partial cross section, Eq.~\eqref{sigJQM}, as in the classical
treatment.

In spite of the similarities between the classical $\sigma_r(\theta,J)$
(Eq.~\eqref{Legdef} or Eq.~\eqref{Gdeflec}) and the quantum $Q_r(\theta,J)$
(Eq.~\eqref{QJ}) there are important differences and hence the qualifier ``analogue''.
The latter is not a genuine  joint probability distribution (and, hence, a true
deflection function in the classical sense) since it includes coherences between
different values of $J$. Moreover, it can take negative values whenever there are
destructive interferences between pairs of $J$ values, although when summed over $J$ a
positive value is recovered. Notwithstanding the differences, as it will be shown in
Section~\ref{Results}, when the interferences are not significant, classical deflection
functions and QM-DF bear a close resemblance.

It is sometimes useful to calculate the angular distributions for a subset of
partial waves These angular distributions, labeled as DCS($J_k$-$J_i$) can be
calculated by restricting the sum in Eq.~\eqref{scattampl} to a given range of
$J$, $J \in [J_i,J_k]$,
\begin{equation}\label{partDCS}
\sigma_r(\theta; J_k - J_i) = \sum_{J=J_i}^{J_k} \sigma_r(J, \theta)
\end{equation}
The partially summed DCS, $\sigma_r(\theta; J_k-J_i)$, include all coherences
between partial waves {\em within} the $[J_i,J_k]$ range but none outside this
range. In addition, like the DCS itself, $\sigma_r(\theta; J_k-J_i)$ are not
additive, especially if there are interferences between different groups of
$J$s.

By analogy, it is also possible to define a deflection function by restricting the sum
over a given  $[J_i,J_k]$ range of $J$, $Q_r(\theta; J_k-J_i)$, as
\begin{equation}\label{sumQJ}
Q_r(\theta; J_k - J_i) = \sum_{J=J_i}^{J_k} Q_r (J, \theta) \, ,\quad \text{with}
~~ J_i \le J_k
\end{equation}
In spite of the similarities between the partial $\sigma_r(\theta; J_k-J_i)$
and $Q_r(\theta; J_k - J_i)$ (and the fact that in the limit of the full
interval, $J_i=0$ and $J_k=J_{\max}$, both functions are identical) there are
two main differences between them: i) The latter also includes coherences between
partial waves outside the $[J_i,J_k]$ range so it may take negative values (if
destructive interferences prevail for some scattering angles); ii)  the
deflection functions so defined, as in the classical case, are additive. Hence, from
the comparison between the partially summed DCSs and partial QM-DFs it is
easy to disentangle the presence and position of interference phenomena.

\subsection{Other Quantum deflection functions}

The idea of a semiclassical deflection function was first developed by Ford and Wheeler
in the context of elastic scattering using the stationary phase
approximation,\cite{Ford-Wheeler:AP59} and later consolidated by Bernstein.
\cite{Bernstein:1966} The semiclassical approximation techniques proved to be very useful
to gain insight into the the physical nature of scattering, making possible to extract
qualitative inferences and easing the interpretation of the quantum results.
\cite{Bernstein:1966,Childbook:1996,Connor:NASI1979}

The semiclassical deflection function, $\Theta(\ell_{\theta})$,  is related to the phase
shift, $\eta_{\ell}$ by
\begin{equation}
\Theta(\ell_{\theta})= 2\, \left(\frac{{\rm d} \eta_{\ell}}{{\rm d}
l}\right)_{\ell_{\theta}}
\end{equation}
where $\Theta =\pm \theta$ for repulsive and attractive potentials,
respectively, and the derivative of $\eta_l$ is evaluated at $\ell_{\theta}$,
the $\ell$-value of the stationary phase. The phase shift can be written in
terms of the $S$ matrix as
\begin{equation}
S_{\ell}= \mathrm{e}^{2 i \eta_{\ell}}
\end{equation}
hence,
\begin{equation}
\Theta(\ell)=  \frac{{\rm d} }{{\rm d} \ell}\left[\arg S_{\ell}\right]
\end{equation}
In a series of articles, Connor and co-workers extended the semiclassical
treatment and developed a quantal version of the deflection function applicable
to the most general case of inelastic or reactive
scattering.\cite{C:PCCP04,SC:PCCP11,XC:JPCA09} It is thus pertinent to compare
our proposed QM-DF with that presented by Connor and coworkers (hereinafter denoted as CQDF). We have
followed the procedure expounded in Ref.~\citenum{C:PCCP04}. In what follows,
we will briefly summarize the main equations of that method for our present
purposes.

For a given initial and final rovibrational states the CQDF, denoted as
$\tilde{\Theta}_{\Omega' \Omega}$, is defined as
\begin{equation}
\tilde{\Theta}_{\Omega' \Omega} (J) = \frac{\,{\rm d}}{{\rm d}\,J} [\arg
\tilde{S}_{\Omega' \Omega}(J)] \, ,
\end{equation}
where  $\tilde{S}_{\Omega' \Omega}(J)$ is the modified scattering matrix
elements that can be calculated directly from the scattering matrix:
\begin{equation}
\tilde{S}_{\Omega' \Omega} (J) = \exp( \imath \pi J) S^J_{\Omega' \Omega}
\end{equation}
It should be highlighted that $\arg \tilde{S}_{\Omega' \Omega} (J)$ does not
denote the principal value, but it is defined as a continuous function as
follows:
\begin{equation}
 \arg \tilde{S}_{\Omega' \Omega} (J) = \arctan\frac{{\rm Im}[\tilde{S}_{\Omega' \Omega}]}
 {{\rm Re}[(\tilde{S}_{\Omega' \Omega}]} + 2 n \pi
\end{equation}
%
\begin{figure}
\centering
  \includegraphics[width=0.9\linewidth]{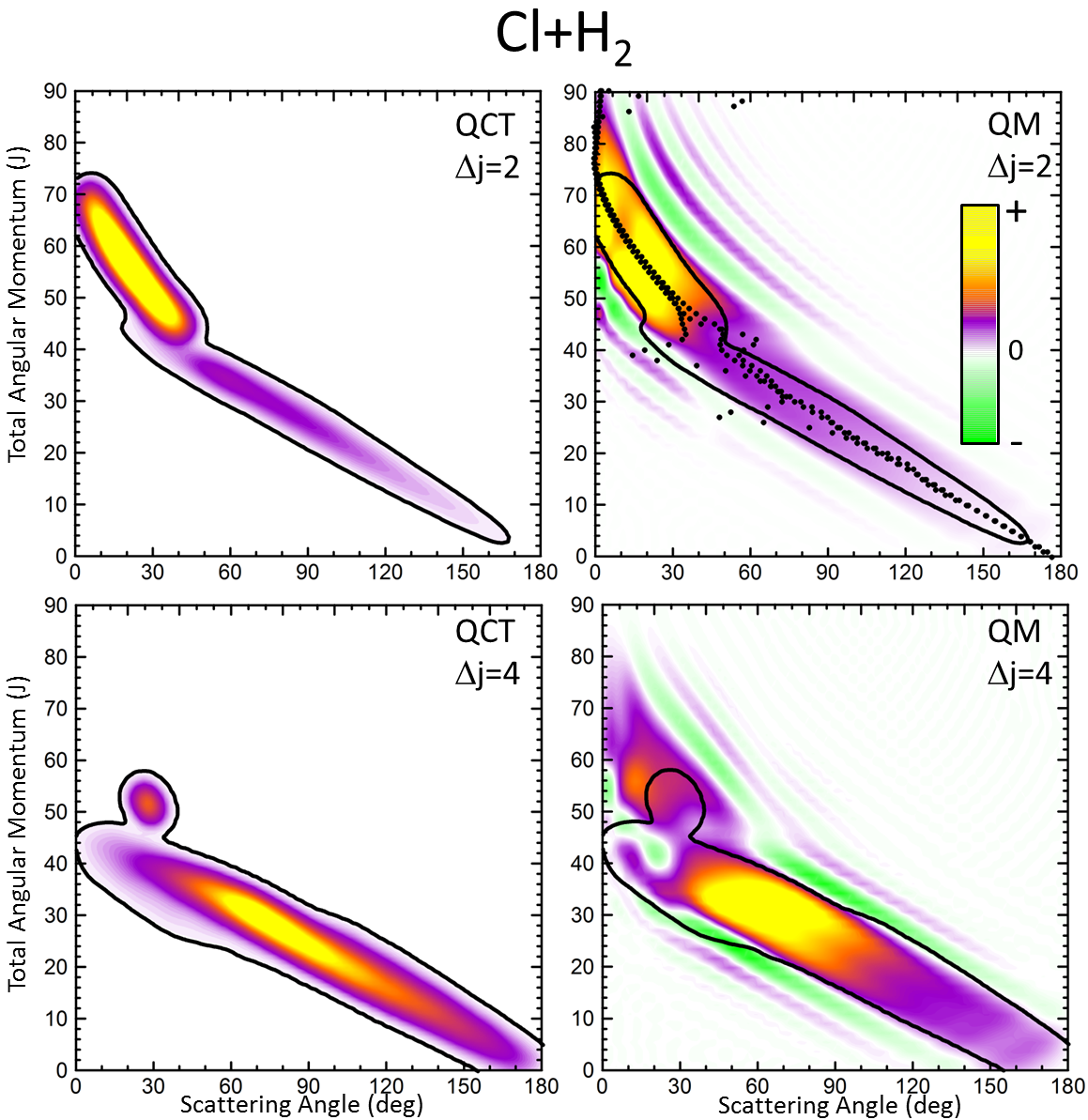}
\caption{Comparison of the QCT deflection functions (left panels) and the
analogue QM deflection function (right panels) for the Cl+H$_2$($v=0$, $j=0$)
$\to$ Cl +H$_2$($v'=0$, $j'=2,4$) inelastic collisions at $E_{\rm col}=$0.73
eV. Top panels, $\Delta j$=2; bottom panels, $\Delta j$=4. The contour of the
QCT deflection function has been added to the QM $Q_r$ to make the comparison
easier. The green colour corresponds to negative value, hence destructive
interferences ($Q_r(\theta, J)< 0$). For comparison purposes, Connor's QDF is
shown on top of the QM analogue deflection function for $\Delta$j=2 using black
dots.} \label{Fig1}
\end{figure}
\begin{figure*}
\centering
  \includegraphics[width=1.0\linewidth]{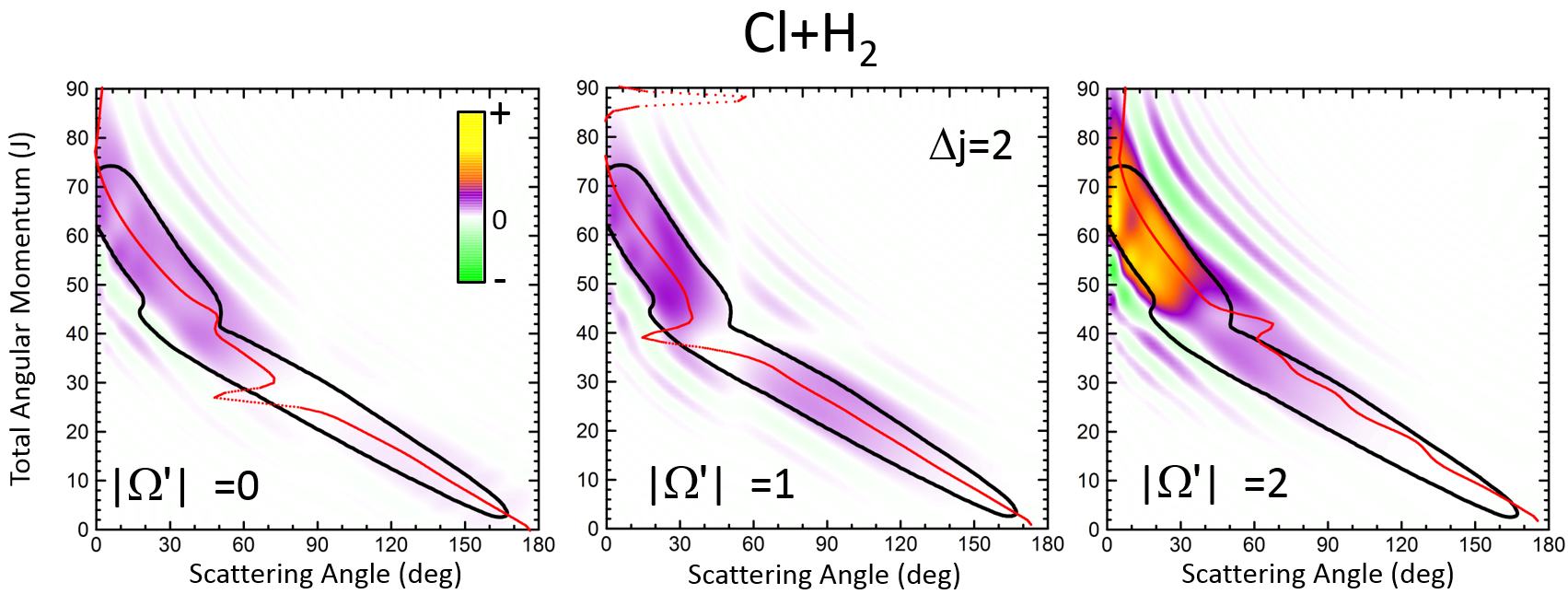}
\caption{QM deflection function at $E_{\rm col}=$0.73 eV for the
Cl+H$_2$($v=0$, $j=0$) $\to$ Cl +H$_2$($v'=0$, $j'=2$, $|\Omega'|=0,1,2$)
inelastic collisions resolved in $\Omega'$ helicity states. The corresponding
Connor's QDF are also shown using solid red lines.} \label{Fig2}
\end{figure*}
where $n$ is a positive or negative integer number, whose value is arbitrarily
set to 0 for $J$=0, and for $J>$0 is selected such that $ \arg
\tilde{S}_{\Omega' \Omega} (J) - \arg \tilde{S}_{\Omega' \Omega} (J-1) < \pi$
is a continuous function. It should be emphasised that whilst $Q_r$ is a sort
of a probability density function in terms of both $\theta$ and $J$, and
therefore contains  information about the scattering intensity and the
presence of constructive or destructive interferences, CQDF
represents a relation between the deflection angle (or the scattering angle)
and the angular momentum $J$. Moreover, as shown in the previous subsection, if
the present QM-DF is summed over over $J$, one gets the DCS.  Another
difference is that whilst CQDF is defined for each pair of $\Omega$ and
$\Omega'$ values, the QM-DF defined in this work can include the average over the
reactant's and the summation over product's helicities as shown in
Eq.~\eqref{QJ}, although it can be also calculated for specific values of
$\Omega$ and $\Omega'$, a it will be shown below.  Apart from these
differences, one would expect a confluence with regard to the relationship
between scattering angle and angular momentum.

\section{Results and Discussion}
\label{Results}

\subsection{Inelastic collisions between Cl and H$_2$}

\begin{figure*}
\centering
  \includegraphics[width=1.0\linewidth]{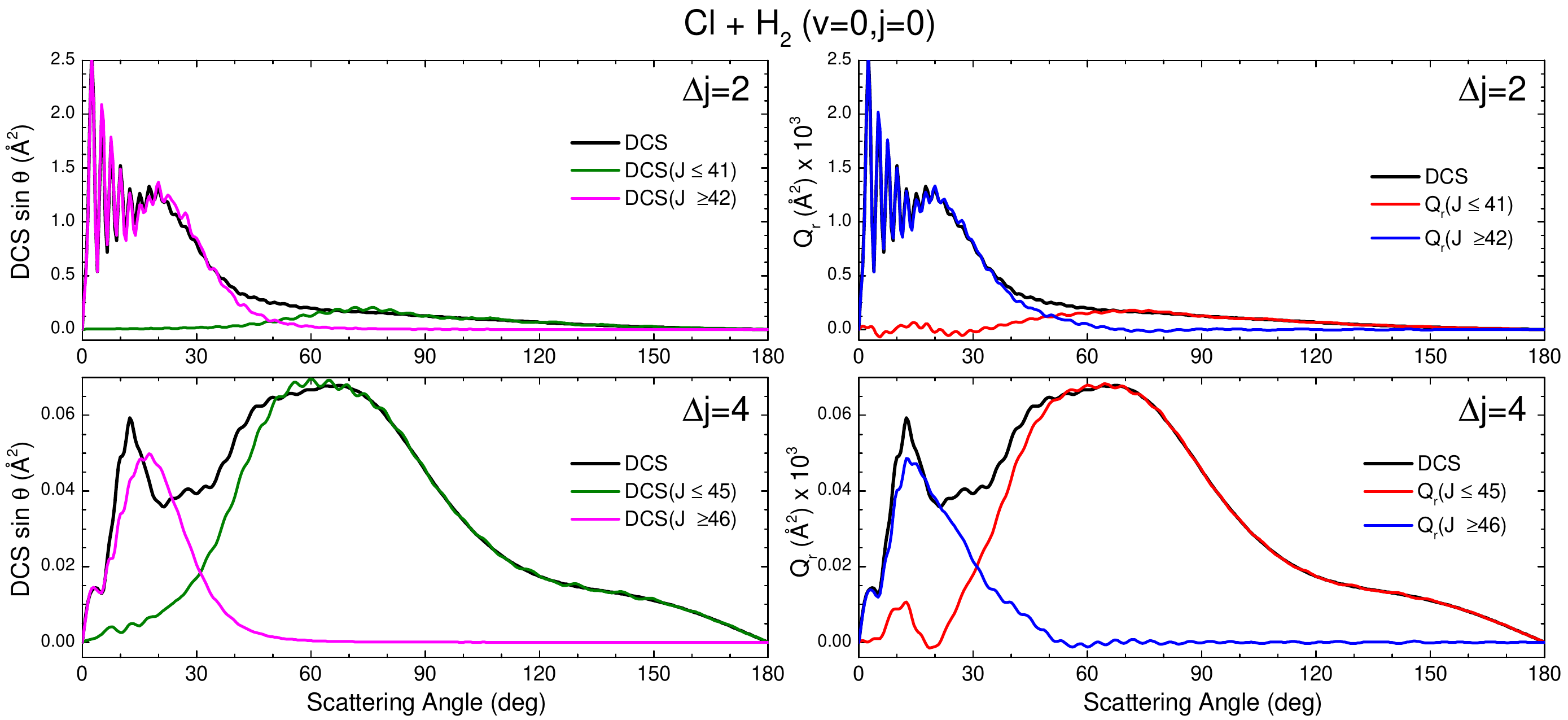}
\caption{Comparison of the DCS partially summed over the indicated $J$
interval, $\sigma(\theta; J_k-J_i)$ (defined in Eq.~\eqref{partDCS}), (left
panels) and the QM deflection functions summed over the same $J$ intervals
$Q_r(\theta;J_k-J_i)$ (defined in Eq.~\eqref{sumQJ})(right panels) for the
inelastic collisions between Cl and H$_2(v=0,j=0)$ at $E_{\rm col}=$0.73 eV and
$\Delta j$=2 (top panels) and $\Delta j$=4 (bottom panels).} \label{Fig3}
\end{figure*}

The first example in which we will use the QM-DF proposed in this work is the
inelastic collisions between Cl and H$_2(v=0, j=0)$. This system has been extensively
studied both computacional and experimentally,
\cite{AC:ARPC03,ABCCVVABSAMT:S96,C:RPP00,WDXCRDWCYJXSLZWA:S08,BSCCSCACW:PRL03} especially
with regard to the role played by the spin-orbit interaction and non-adiabatic effects
for the hydrogen exchange reaction.

As for inelastic collisions, some interesting features emerged in previous
studies.\cite{GAJA:JCP11,AAGJMS:PCCP12} QM and QCT calculations using the BW2 PES
\cite{BW:JCP00} showed that at relatively high collision energies ($E_{\rm coll}>0.6$ eV)
and for small $\Delta j$ values ($\Delta j= j'-j$), the inelastic probabilities,
$P_r(J)$, exhibit two maxima separated by a minimum in the QCT and QM results. This
minimum was identified as that corresponding to the glory impact parameter. The analysis
of the results showed that there are two mechanisms responsible of the inelastic
scattering resulting in very different stereodynamical behaviours, and associated to
different regions of the PES.\cite{GAJA:JCP11,AAGJMS:PCCP12} The two distinct dynamical
regimes depend primarily on the value of the total (here also orbital) angular momentum:
(i) for $J$s below the glory impact parameter, collisions seem to take place following a
sort of ``tug-of-war'' mechanism \cite{GWGZMZ:N08} that implies the stretching of the H-H
bond,\cite{AAGJMS:PCCP12}; and (ii) for $J \gtrsim 40$ collisions can be assigned to
rainbow scattering in which the attractive part of the PES is sampled.\cite{GAJA:JCP11}
For transitions implying higher $\Delta j$, that require more head-on collisions, the
contribution of high impact parameters wanes rapidly, and the second maximum in the
$P_r(J)$ leading to small scattering angles disappears. The semi-quantitative agreement
between the classical and quantum $P_r(J)$ and DCSs seems to indicate that quantum
effects associated to interferences between the two groups of partial waves are not
expected to be important.\cite{GAJA:JCP11} Therefore, the Cl+H$_2(v=0, j=0)$ inelastic
scattering seems to be a good example of a collision system in which the QM-DF as
proposed in this work would resemble the QCT deflection function.

Figure~\ref{Fig1} displays the QCT and the QM deflection functions the $j=0 \to j'=2$ and
$j=0 \to j=4$ transitions (top and bottom panels, respectively) at $E_{\rm coll}=$0.73
eV. The l.h.s panels show the QCT $\sigma(\theta,J)$.  The two different dynamical
regimes can be easily distinguished. For $\Delta j$=2, the high$-J$ mechanism is
preeminent and gives rise to scattering into $\theta <50^{\circ}$. The low-$J$ mechanism
appears in the deflection function as a narrow band that extends from
$\theta$=40$^{\circ}$ to $\theta$=180$^{\circ}$ and comprises $J$ values from 0 to 40.
The negative slope, common to both regimes (although with different values) is
characteristic of direct collisions and follows the simple correlation of low (high)
impact parameters leading to high (small) scattering angle. For $\Delta j$=4, the
prevailing mechanisms is that corresponding to $J \le 40$ values, and the high-$J$
mechanism appears as an small island in the $\theta-J$ map, centered at $J=50$ and
$\theta=$30$^\circ$.

The equivalent QM $Q_r(\theta, J)$'s, shown in the right panels of Fig.~\ref{Fig1}, bear
close similarities with their classical counterparts, although with some noticeable
differences. For $\Delta j$=2, the high-$J$ mechanism, responsible of most of the
scattering, extends to larger values of $J$, it is also broader, and it is flanked by a
series of stripes, some of negative value (green colour) associated to destructive
interferences. The negative slope of the low-$J$ mechanism is also observed, although in
this case both mechanisms merge at $J\sim$ 45. There are also a series of negative
stripes parallel to the main band that cause a small decrease of the DCS. It should be
noticed that, for the sake of clarity in the figure, the QM-DF have been smoothed given
the discrete character of $J$. The same procedure will  be followed for all remaining 3D
plots of this article. For $\Delta j$=4, the QM-DF also extends to larger $J$ values and
the high-$J$ mechanism covers a broader $J-\theta$ region than in the QCT case. As in the
classical case, for this transition, the low-$J$ mechanism bears away most of the
scattering.

\begin{figure*}
\centering
  \includegraphics[width=0.9\linewidth]{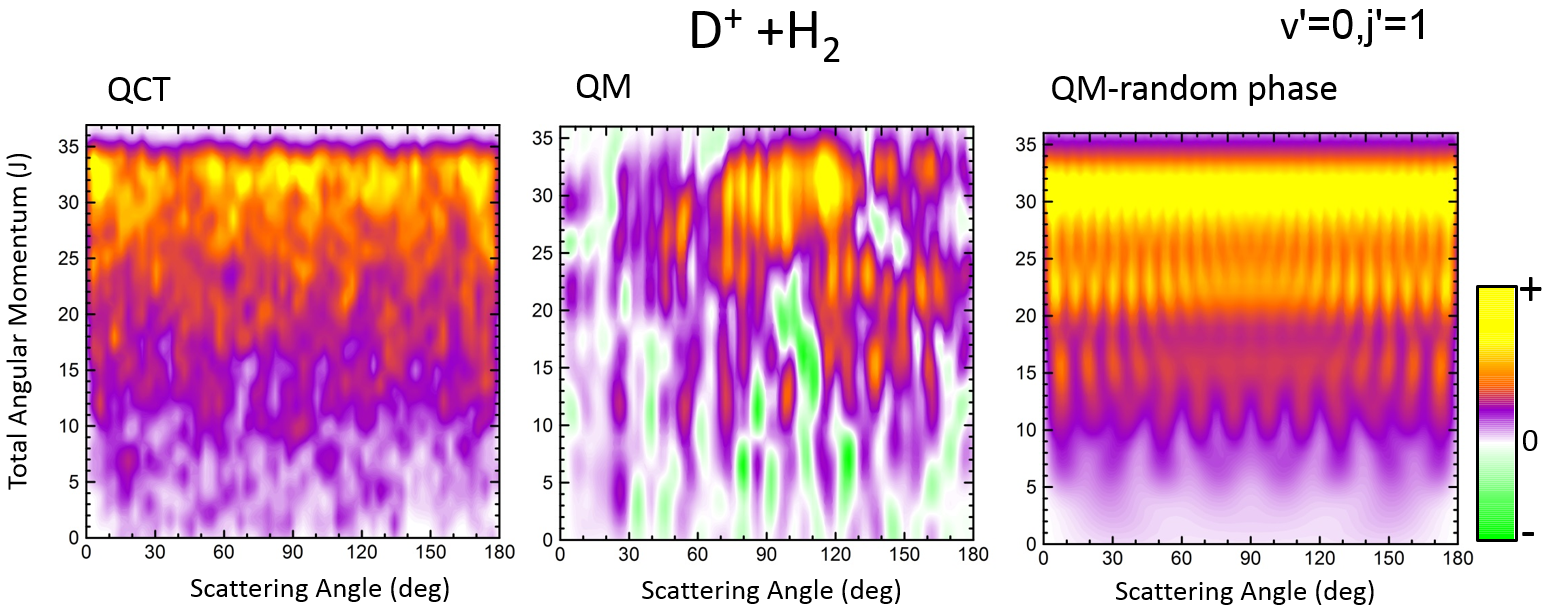}
\caption{Comparison of the QCT deflection function and the QM analogue
deflection function for D$^+$ + H$_2$ reaction at $E_{\rm col}$=0.15 eV. The
results are  for HD($v'$=0, $j'$=1).} \label{Fig4}
\end{figure*}

The results of the CQDF for $\Delta j$=2 are also shown as a dotted lines along with the
present $Q_r(\theta, J)$. The points corresponding to $\Omega'=$0, 1 and 2 are all
included. As can be seen, the $\tilde{\Theta}_{\Omega' \Omega}(J)$ follows almost exactly
the middle line (reproducing the two different slopes) of the present QM-DF and is also
in good agrement with the QCT deflection function. More detailed information is shown in
Fig.~\ref{Fig2}, where the $Q_r(\theta, J, \Omega')$ are plotted separately for each of
the three possible $\Omega'$ values along with the corresponding CQDF,
$\tilde{\Theta}_{\Omega' \Omega=0}(J)$. As can be seen, the agreement is excellent and
CQDF matches almost exactly the most probable dependence of $\theta$ with $J$ found with
the present QM-DF. It should be pointed out, however, that the latter carries information
on the intensity of scattering for each $J-\theta$ region, and about the presence of
constructive and destructive interferences. Indeed, the information conveyed by the
present $Q_r(\theta, J, \Omega')$ goes well beyond that obtained by the  CQDF. As can be
seen, most of the intensity of the high-$J$ mechanism corresponds to $\Omega'$=2,
indicating that the product's $\bm j'$ rotational angular momentum lies preferentially
along the recoil velocity, whilst that corresponding to low-$J$ is more isotropic with
some preference for $\Omega'$=1.\cite{AAGJMS:PCCP12}

The partial DCS, Eq.~\eqref{partDCS}, and the QM-DF summed over the indicated
range of $J$, Eq.~\eqref{sumQJ}, are shown in left and right panels of
Fig.~\ref{Fig3} for $\Delta j$=2 and 4, respectively. The two $J$ intervals
have been chosen to comprise partial waves corresponding to the low-$J$ ($J \le
41$ for $\Delta j$=2 and $J\le 45$ for $\Delta j$=4) and high-$J$ ($J > 41$ for
$\Delta j$=2 and $J>45$ for $\Delta j$=4). Therefore, the two magnitudes are
broken down in their contributions from the two intervals for comparison
purposes. It should be recalled that if the whole range of $J$ is included,
both magnitudes become identical, corresponding to the total (converged) DCS.
However, whilst the partial DCS only encompasses those coherences only within
the chosen interval, the partially summed QM-DF comprises all possible coherences
(although their contribution are halved) internal and external to that
interval.

The first consideration to be held is the similarity of the respective decompositions of
the partial DCSs and the summed QM-DFs $Q_r(\theta; \Delta J)$, of the left and right
panels. As a second consideration, for $\Delta j=2$, the incoherent sum of
$\sigma(\theta; J=0-42)$ and $\sigma(\theta; J>42)$ reproduces fairly well the converged
(total) DCS (recall that the partial DCS are not additive), evincing that interferences
between the two mechanisms are practically negligible. A similar analysis was performed
in Ref.~\citenum{AAGJMS:PCCP12} leading to the same conclusion. This is further confirmed
by inspection of the $Q_r(\theta; \Delta J)$, shown in the right-top panel, which
are almost identical to the partial DCSs, except for few differences in the forward
region. For the case of $\Delta j=4$ the situation is much the same as that for $\Delta
j=2$. The only, main difference between partial DCSs and $Q_r(\theta; \Delta J)$ can be
observed at forward scattering angles $\theta=$10$^\circ$--30$^\circ$. As can be seen,
there is a peak centred on $\theta=12^\circ$ in the $Q_r(\theta; J < 46)$ which is absent
in the respective partially summed DCS. This implies that, although without being
substantial, there are still some interferences between the two groups of partial waves.
Returning to Fig.~\ref{Fig1}, it is possible to associate this effect with the feature
that appears with a `hook' at the top corner of the right-bottom panel of that figure.

\begin{figure}[ht!]
\vspace{-1.0cm} \centering
  \includegraphics[width=0.9\linewidth]{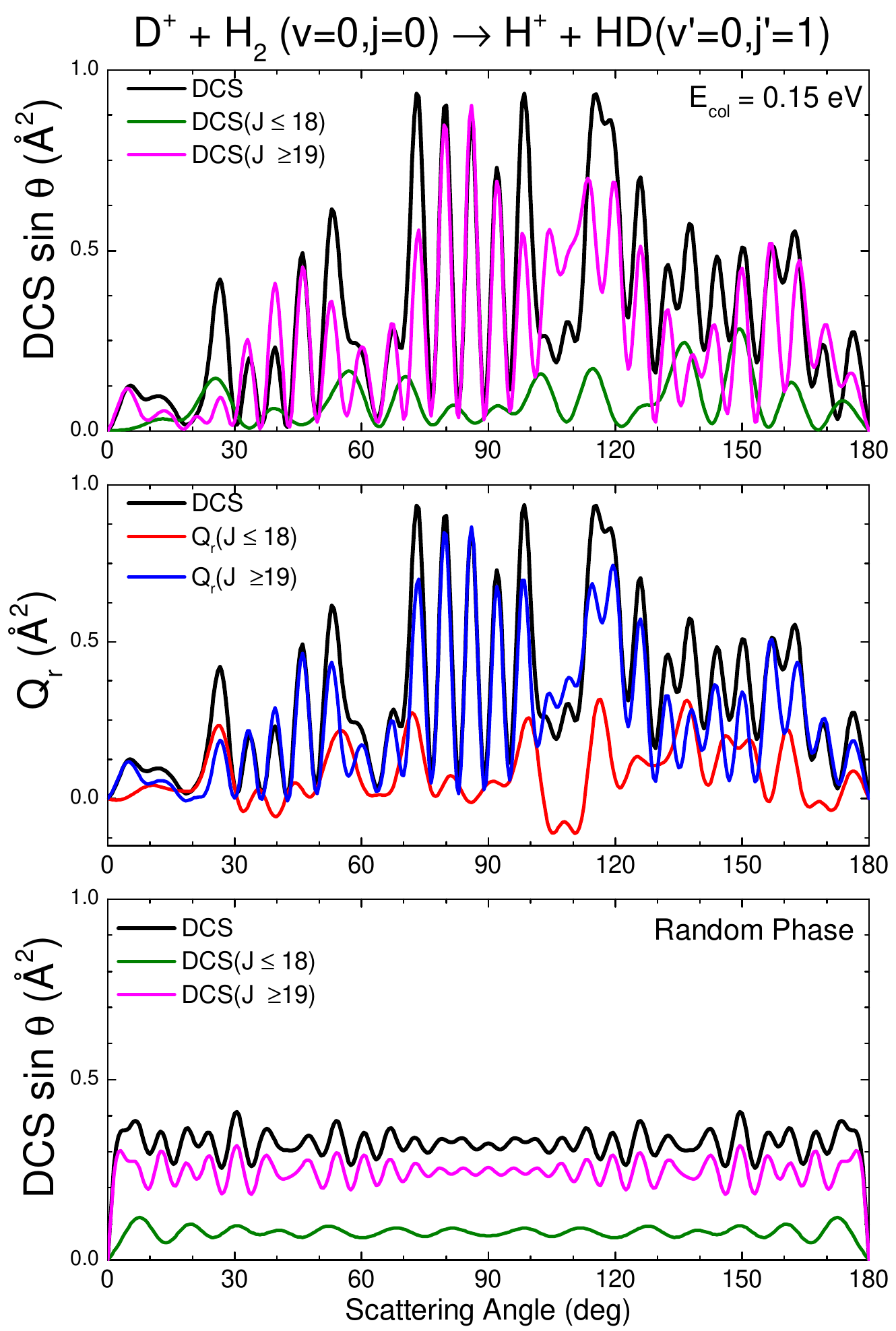}
\caption{Comparison of the $Q_r(J,\theta)$ and the DCS for the D$^+$ + H$_2 \to
$ HD($v'$=0,$j'$=1) + H$^+$ reaction at $E_{\rm col}=$0.15 eV. } \label{Fig5}
\end{figure}

It must be pointed out that the above discussion does not imply that for
$\Delta j=2$ there are not interferences within one of those groups of partial
waves. By inspection of right-bottom panel of Fig.~\ref{Fig1}, it is obvious
that in the forward region and high $J>40$ there are many positive and negative
interferences that are the origin of the oscillations observed at
$\theta<30^{\circ}$ in Fig.~\ref{Fig3}.

\subsection{Reactions that go throw a long-lived complex, D$^+$ + H$_2$}

A contrasting system is the D$^+$ + H$_2 \to$HD+ H$^+$ reaction on its first 1$^1A'$
adiabatic PES. As is well known, this PES is barrierless and rather featureless,
overwhelmingly dominated by a very deep well  of 4 eV from the
asymptotes.\cite{ARTSP:JCP00,VLABP:JCP08} Given its importance in astrochemistry, it has
been extensively studied both theoretical and experimentally (see, for example,
\citenum{G:JCSFT93,G:PS95,GH:CR92,GS:PSS02,GH:MNRAS17,JAEHR:JCP09,JAGHHSA:PCCP12,JABSBH:PCCP10,JAAHS:PCCP10,ZRGRASG:JPCA09,LCH:JPCA05}
and references therein). It has been long assumed that given the absence of barrier and
the presence of a deep well, the H$^+$+H$_2$ reaction could be considered as a prototype
of statistical reaction. However, although at low collision energies quantum,
quasiclassical and statistical approaches seem to converge yielding results essentially
coincident (apart from rapid oscillations), at higher energies, that imply large
values of $J$, the centrifugal barrier tends to wash out the potential well potential
and QCT calculations indicate that the residence times in the well are too short for the
reaction to behave statistically.
\cite{JAEHR:JCP09,JAGHHSA:PCCP12,JABSBH:PCCP10,JAAHS:PCCP10}

We will focus on the results at a sufficiently low energy, $E_{\rm coll}=$ 150 meV and
HD($v'$=0,$j'$=1) formation, where the statistical (ergodic) assumption seems to hold.
Indeed, at that energy, the D$^+$ + H$_2$ reaction proceeds through formation of a
long-lived complex,  the shape of $P(J)$ and the product state distributions follow the
trend predicted by statistical methods.\cite{JAAHS:PCCP10} Hence, this seems to be a good
example to test the reliability of the $Q_r(\theta,J)$ in statistical reactions. In
Fig.~\ref{Fig4} three deflection functions are shown: the classical deflection function,
the QM-DF and the quantal one under the assumption of the random phase approximation,
which assumes that there is not correlation between different $J$s, so that a
deflection function equivalent to the classical one can be calculated.  In all three
cases, as expected for a statistical reaction, there is no clear correlation between $J$
and $\theta$: all $J$ seems to contribute to every scattering angle. The only remarkable
feature in the classical deflection function is the largest probabilities found at high
$J$, due to the fact that the $P_r(J)$ is flat until it decreases abruptly when reaching
$J_{\rm max}$. The $Q_r(\theta,J)$, shown in the right panel of Fig.~\ref{Fig4},
indicates the presence of many destructive (green) and constructive (red/yellow)
interferences that will give rise to multiple oscillations in the DCS over the whole
range of scattering angles. However, coherences even if they occurred between partial
waves with separated $J$ values, are so numerous that their effect is smoothed out to
some extent. This is the basic assumption in the random phase
approximation,\cite{WL:JCP71,RHM:CPL01} that allows one to calculate coarse-grained
product's state distributions DCSs and other vector correlations \cite{JAMSA:PCCP12} by
neglecting the coherences between different total angular momenta, and hence with a
formidable saving of computational effort. The right panel of Fig.~\ref{Fig3} shows the
random phase approximated DF, where all the coherences have been neglected by only
keeping the diagonal term, $|f^J_{\Omega' \Omega}|^2$, in Eq.~\eqref{QJ2}. Apart from the
discrete character of $J$, the similitude with the QCT deflection function is remarkable.
For this reaction we do not show the results obtained using the CQDF as a
single valuated function per $J$ cannot account for the complex  pattern depicted in
Fig.~~\ref{Fig4}. For this reaction, CQDF results in a highly oscillating function due to
the superimposed of nearside and farside scattering \cite{HC:AIP15}.

The partial and total DCSs, as well as the QM-DF summed over limited ranges of
$J$, are shown in the top and middle panel, respectively, of Fig.~\ref{Fig5}.
The $J$ dividing value between low-$J$ and high-$J$ values, has been chosen
somewhat arbitrarily as $J_{\max}/2$, since not hint of change of mechanism can
be appreciated in neither the QCT nor QM-DF.  In the latter case, since no
coherences are considered, both magnitudes given by Eqns.~\eqref{partDCS} and
\eqref{sumQJ} are identical as only the $|f^J_{\Omega' \Omega}|^2$ terms are
included. As expected from the QM-DF, the DCSs with the full QM calculation
exhibit many oscillations in the whole range of scattering angles, reflecting
the numerous interferences that were apparent in Fig.~\ref{Fig4}. The partial
DCSs and their respective $Q(\theta, \Delta J)$ summed in the $[0,18]$ and
$[19,35]$ are fairly similar. If we recall that the former are only sensitive
to those coherences within the chosen interval whereas the latter includes all
of them inside and outside the chosen interval, one can conclude that
interferences between separate $J$ values exist but, overall, they almost cancel out.

The partial DCSs, which under the random phase approximation coincides with the
$Q_r(\theta, \Delta J)$ (summed over $J$) is shown in the bottom panel of
Fig.~\ref{Fig5}, still show some oscillations, nothing surprising considering that they
are basically the result of the summation of $[d^J_{\Omega \Omega'}(\theta)]^2$ terms,
two for each partial wave. In any case, they correspond, as expected, to the average QM
DCSs in which the oscillations have been washed out. The resulting random phase DCSs are
strictly symmetric, peaking at forward and backward angles (recall that the represented
DCSs have been multiplied by $\sin \theta$). Although at first glance there seems to be a
poor approximation of the actual DCSs, but it must be borne in mind that the
observed oscillations change rapidly with the collision energy and initial states, hence
they would be barely discernible under experimental conditions.

\subsection{Direct Reactions: H + D$_2$}
\begin{figure}
\centering
  \includegraphics[width=1.1\linewidth]{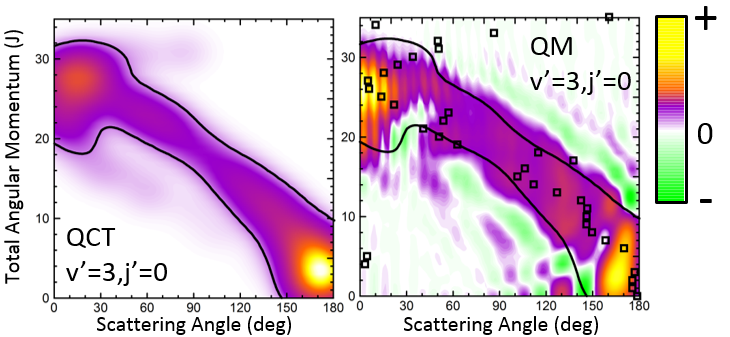}
\caption{QCT and QM deflection functions for the H + D$_2 \to$ HD($v'$=3,
$j'$=0)+D reaction at $E_{\rm col}=$1.73\,eV. The contours of the classical deflection functions
are copied in the plots representing the QM-DFs to highlight the similarities
and differences. The black squares represent the Connor's $\Theta_{0 \,0}(J)$.}
\label{Fig6}
\end{figure}

The third system we will be concerned with is the H + D$_2$ reaction, possibly the most
widely studied reaction, and indeed the benchmark system in reaction dynamics. Although
from many points of view can be considered as the simplest reaction, its dynamics is far
richer than it could be expected;\cite{ABH:IRPC05,GMW:JCP08,GMWA:JCP08} indeed, when
investigated in detail still renders unexpected
results.\cite{DWHWYCS:S03,JSZBAHA:PNAS14,JHASJZ:NC15} Very recently, the angular
distributions of state resolved HD formed in collisions between H and D$_2$ at $E_{\rm
col} = 1.73$ eV were measured using the photoloc technique.\cite{SOSXZ:CPL93} For
HD($v'$=1,low $j'$) states the angular distributions in the backward hemisphere
were dominated by a series of peaks and dips whose origin was traced to interferences
between the two mechanisms described in Refs.~\citenum{GMWA:JCP08,JHASJZ:NC15}. For both
higher $v'$ and/or $j'$ rovibrational states, one of the mechanisms disappears and so
does the interference pattern in the DCS. In previous works it was shown that the QCT
deflection functions was crucial for the right interpretation and assignment of the
observed interference pattern.\cite{JHASJZ:NC15,JAASZ:CS16} It can be thus expected that
the QM-DF will carry at least the same and presumably even more information about the
mechanism. Therefore, the state resolved H + D$_2$ reaction would be an excellent system
to test the quantum analogue to the classical deflection function as we can test its
performance under three different scenarios: (i) HD($v'$=1,$j'$=0) formation, where the
interference pattern is conspicuous and dominates the shape of the DCS in the backward
hemisphere; (ii) higher $j'$, for instance HD($v'$=1,$j'$=5),  where oscillations start
to disappear; (iii) higher $v'$, {\em v.g.}, HD($v'$=3,$j'$=0), where no oscillations
were observed in the DCS. In what follows we will show the QM-DF, partial DCS and the
QM-DF summed over the appropriate ranges of $J$ for these three different scenarios. All
calculations were carried out on the BKMP2 PES.\cite{BKMP:JCP96}
\begin{figure}
\centering
  \includegraphics[width=0.80\linewidth]{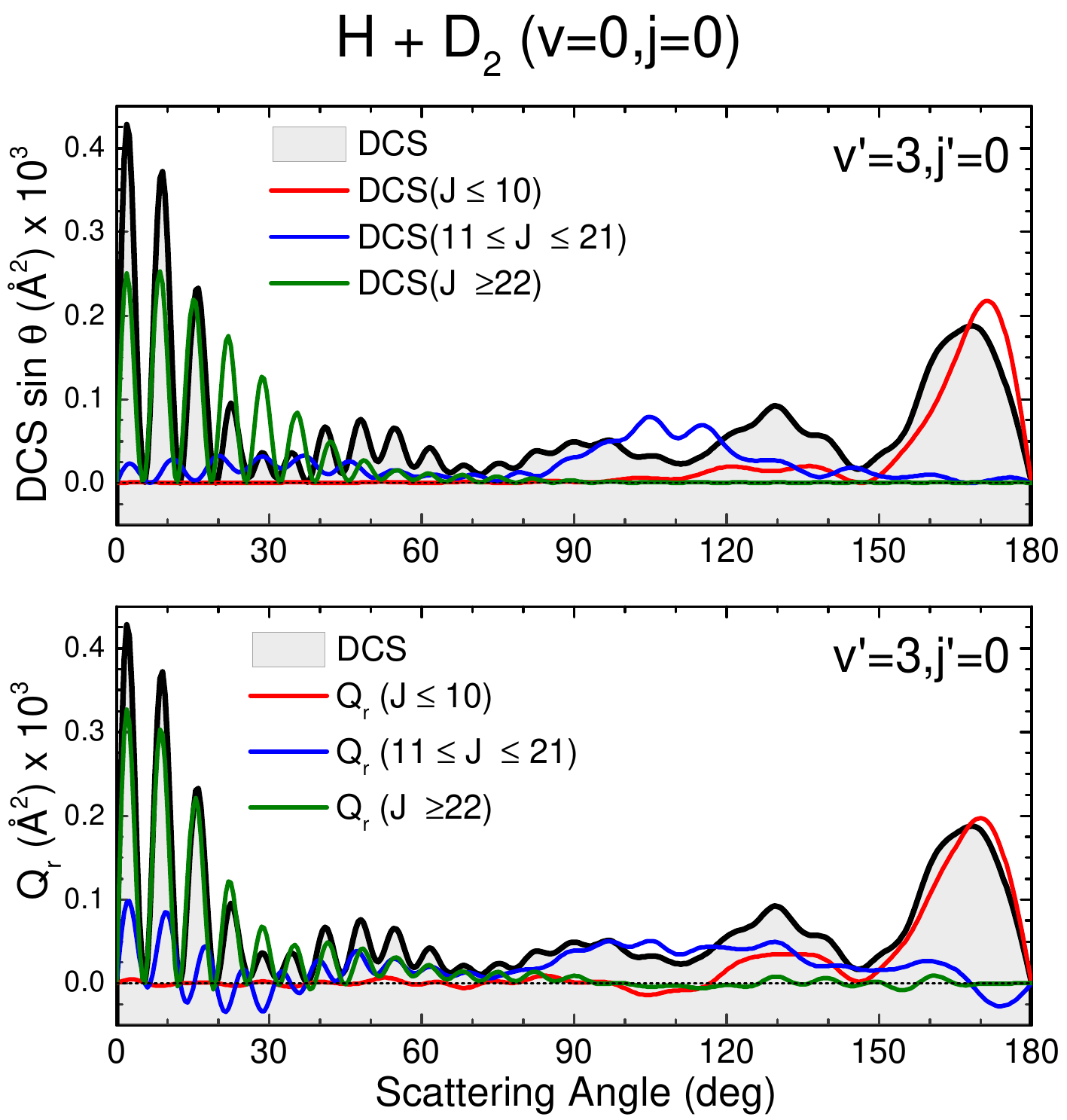}
\caption{Comparison of the partial DCS (upper panel) and the QM-DF summed over
the same $J$ contributions, $Q_r(\theta, \Delta J)$ (bottom panel) for the H +
D$_2 \to $HD($v'$=3,$j'$=0)+D reaction at $E_{\rm col}=$1.97 eV.} \label{Fig7}
\end{figure}
\begin{figure}
\centering
  \includegraphics[width=1.0\linewidth]{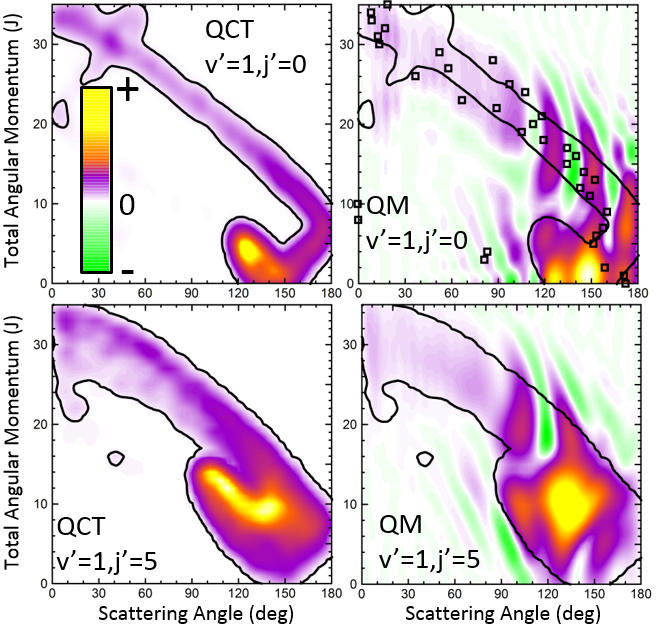}
\caption{QCT (left) and QM (right) deflection functions for the H + D$_2 \to$
HD($v'$=0, $j'=0, 5$)+D reaction at $E_{\rm col}=$1.73\,eV. The results for
$j'$=0 and $j'$=5 are shown in the top and bottom panels, respectively. The
contours of the classical deflection functions are copied in the plots representing the QM-DFs
to highlight the similarities and differences. For the HD($v'$=0, $j'$=0)
formation, Connor's $\Theta_{0 \,0}(J)$ is also represented as black squares.}
\label{Fig8}
\end{figure}
%

Let us first turn our attention to those collisions leading to
HD($v'$=3,$j'$=0) whose classical deflection function and QM-DF are depicted in
Fig.~\ref{Fig6}. The QCT deflection function shows the typical profile for a
direct reaction mechanism, similar to that observed for the inelastic
collisions between Cl and H$_2$, {\em i.e.}, a band running diagonally across
the $\theta$-$J$ map, with low $J$ giving rise to backward scattering and high
$J$ correlating with forward scattering. In this case, the mechanism covers the
whole range of scattering angles with one maximum in the forward and another in
the backward region. Moreover, there seems to be no other mechanism to compete
with it. Not surprisingly, QCT and QM-DFs are very similar, showing the same
structure moving from backwards to forwards. However, although the quantum
results were somewhat smoothed out for the sake of clarity, we can still
observed series of constructive and destructive interferences manifested as
stripes, especially in the forward scattering region. In addition, the main
band is flanked by two green stripes (destructive interferences) that will give
rise to oscillations in the DCS.

Figure~\ref{Fig7} depicts the partial DCS and the QM-DF summed over three
subsets of partial waves that, according to the deflection functions of Fig.~\ref{Fig6}, can be
associated to backward ($J \in [0,10]$), sideways ($J \in [11,21]$) and forward
($J>22$) scattering. There is a remarkable similitude between the partial DCSs
and $Q(\theta, \Delta J)$ for each of the three subsets of $J$ used in the
decomposition of these magnitudes, implying that there are essentially  no
interferences between the partial waves belonging to different subsets. Only at
forward scattering angles there are some appreciable interferences between
partial waves associated to $J$ values of  $J \in [11,21]$ and $J>22$ subsets.
There is one more aspect that deserves a comment. The maxima and minima that
can be observed in the DCS can be easily inferred from the positive and
negative values of the QM-DF. In particular, the minima at 70$^{\circ}$,
115$^{\circ}$ and 150$^{\circ}$ correspond to the negative (green colour)
contributions in the QM-DF. These minima (and the precedent or subsequent
maxima) cannot be deduced from the classical deflection function.
\begin{figure*} [ht!]
  \includegraphics[width=0.80\linewidth]{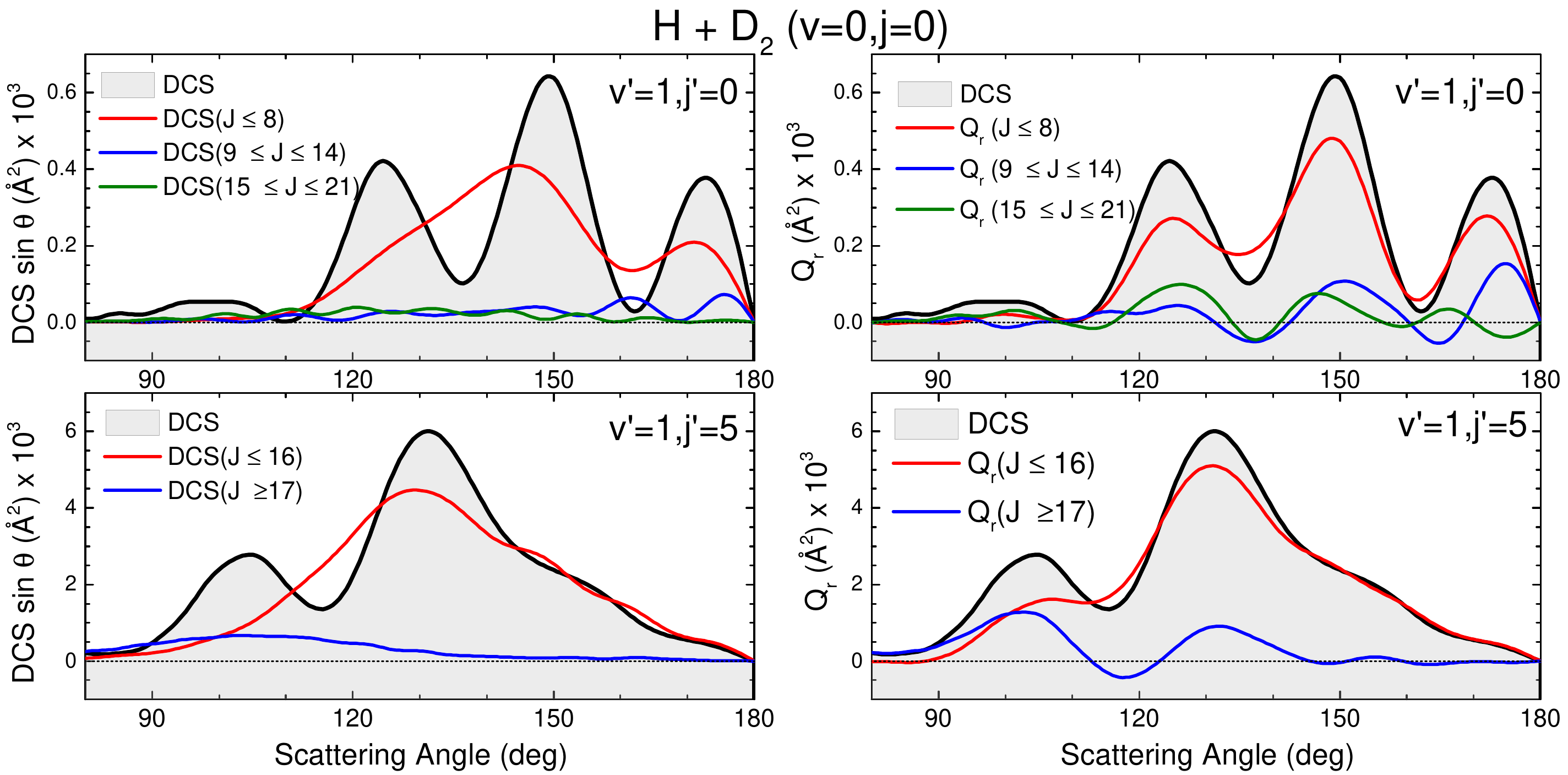}
\caption{Comparison of the partial DCS (left panels) and $Q_r(\theta, \delta
J)$ (right panels) for the H + D$_2$ reaction at $E_{\rm col}=$1.97 eV. Top
panels is for scattering giving rise to HD($v'$=1,$j'$=0), whilst the bottom
panel corresponds to $v'$=1,$j'$=5. Only the backward hemisphere is shown for
clarity.} \label{Fig9}
\end{figure*}

Let us now move to the collisions leading to HD($v'$=1,$j'$=0). The QCT and QM-DFs are
shown in the top panels of Fig.~\ref{Fig8}. As it was discussed at length in previous
work \cite{JHASJZ:NC15}, and can be seen by inspection of the QCT deflection function,
there are two main, distinct mechanisms that are likely to interact with each other
giving rise to the interference pattern observed experimentally. One of them corresponds
to the main band with a negative slope, similar to that we have found for $v'$=3; the
other mechanism, confined in a reduced region of the $J$-$\theta$ map, between
110$^{\circ}$-160$^{\circ}$ and low $J$ values, accounts for most of the reactivity.
Between them, as a sort of bridge, there is still a third mechanism with a positive slope
that comprises low values of $J$ and $\theta >160^{\circ}$. Using the QCT deflection
function it easy to predict that interferences will take place,\cite{JHASJZ:NC15} since
different paths are leading to the same scattering angles. However, the classical
deflection function cannot predict the interference pattern: how many oscillations and
what would be their positions. In previous examples, we have shown that the QM-DFs were
akin to their QCT counterparts. Admittedly, we could gain some additional information
from the formers, but the gist of the processes could be captured by the classical
deflection functions. In this example, however, we will see that the quantum $Q_r(\theta,
J)$ provides an additional and most valuable information.

The first observation is that the QM-DF shown in the top-right panel of Fig.~\ref{Fig8}
is rather different to its classical counterpart. Only with the help of the superimposed
countour of the classical $\sigma_r(\theta, J)$ and leaving aside the destructive
coherences, we could see that they share the main gross features. Even then, the QM-DF is
broader and the region corresponding to the diagonal band almost merges with the
mechanism confined between 110$^{\circ}$-160$^{\circ}$ and $J <10$. But the main source
of discrepancy lies on the presence of negative, destructive (green colour) and positive,
constructive (red colour) interferences that do not flank the main band -- as in the case
of HD($v'$=3,$j'$=0) scattering-- but they are transversal to it, cutting the diagonal
band in several slices. Since $Q_r(\theta, J)$ are additive, it is easy to realize that
each of the slices corresponds to the various peaks in the DCS, whilst the vertical green
stripes corresponds to minima in the DCS. Therefore, just looking at the QM-DF we could
discern: (i) that there will be three peaks in the backward hemisphere, (ii) which will
be their positions, as well as those of the respective minima, and (iii) the partial
waves that contribute to each of the peaks.

Not surprisingly, the partial DCS and the QM-DFs summed over range of $J$ values,
$Q_r(\theta, \Delta J)$, calculated for subsets of partial waves and shown in
Fig.~\ref{Fig9} do not look alike. The DCS($J\le8)$ can be associated to the
confined mechanism and, although it carries most of the reactivity, it shows a
broad, blunt shape with no hint of the three finger-like peaks present in the
total DCS in the 100$^{\circ}$-180$^{\circ}$ range. Clearly, the sole
consideration of coherences within the $J \in [0,8]$ interval, which are the
only ones in the partial DCS, is unable to predict the shape of the DCS. In
stark contrast, the $Q_r(\theta,0\le J \le 8)$, that accounts for all the
coherences in and outside the $J\in[0,8]$ range, looks similar to the overall
DCS. The partial DCSs calculated for $J>8$  ($J \in [9,14]$ and $J \in
[15,21]$) are very small throughout the whole range of scattering angles,
whereas their respective $Q_r(\theta, \Delta J)$ are not that small. On top of
that, at some angles they are negative, a consequence of the negative contour
shown in Fig.~\ref{Fig8}.

The third scenario corresponds to collisions leading to HD($v'$=1,$j'$=5) whose
QCT and QM-DFs are portrayed in the bottom panels of Fig.~\ref{Fig8}. As can be
seen, the structure that was isolated for HD($v'$=1,$j'$=0) has almost merged
into the diagonal band and is considerably less confined. In addition, QCT and
QM-DFs look now more alike. Yet the the main band is cut by the signature of a
destructive interference (the green slice at  $\theta \sim 115^\circ$) that can
be expected to give rise to a minimum in the backward DCS.

The comparison of the partial DCS and the $Q_r(\theta, \Delta J)$ confirm these findings
and clarifies the role of interferences. The choice of $J$=16 for the decomposition seems
to be a sensible choice at the light of the deflection functions shown in
Fig.~\ref{Fig8}. In contrast to the results for HD($v'$=1,$j'$=0), the DCS($J\le 16$) is
similar to $Q_r(\theta, J \le 16)$, although the later is somewhat more structured.
However, the $Q_r(\theta, J \ge 17)$ displays some oscillations and a negative
contribution at $\approx 115^\circ$ (as expected from the green slice commented on above)
that reveals coherences with the low subset of partial waves. The effect of these partial
waves is to sharpen the shape of the DCS, defining more clearly the two maxima and the
intermediate minimum.

Finally, it is worthwhile to compare the results obtained using the formalism devised in
this work with the CQDF. In Figs.~\ref{Fig6} and \ref{Fig8}, superimposed to the
$Q_r(\theta, J)$, the respective CQDF for $v'$=3 $j'$=0 and  $v'$=1 $j'$=0 are
represented as open squares.  In both cases the agreement is fairly good, covering the
regions occupied by the present QM-DF. In particular, the oscillations observed in
extreme forward for $v'$=3, that could be predicted  by the $Q_r(\theta, J)$, can be also
foreseen using CQDFs (different $J$s leading to the same $\theta$).  In fact, using CQDF
it can be concluded that they are caused by interferences between nearside and farside
reactive flux. \cite{SC:JCP12} For the $v'$=1 case, however, the sole analysis of the CQDF barely
accounts for the confined, predominant mechanism. It must be pointed out that even if we
could observe the various mechanisms in the CQDFs, it would have not been possible to
predict neither the number of peaks and dips or their position since, for its
construction, it only provides one single value of the deflection angle per partial wave.

\section{Conclusions}

The joint dependence of scattering intensity on the angular momentum and scattering
angle, represented by the classical deflection function, has proved to be extremely
useful to unravel the mechanism of a colliding system. Indeed, from its inspection one
can disentangle reaction mechanisms as well as allows us to predict the presence of
interferences. However, the classical deflection function is an ill defined concept on
pure quantum mechanical grounds as the differential cross section depends on the
coherences between the different values of $J$ caused by the cross terms in the expansion
of partial waves. In this work we propose a conceptually simple  quantum analogue to the
classical deflection function that does account for the coherences and whose
interpretation is rather intuitive. Moreover, as it has been defined, the quantum
analogue to the classical deflection function (QM-DF) not only  relates scattering angles
with angular momenta but also accounts for the scattering intensity. As such, summing
over the whole set of angular momenta for convergence yields the DCS, and integrating
over scattering gives the reactive (or inelastic) partial cross section, similarly to the
classical deflection function.

Throughout this article we have applied  the proposed QM-DF to several case
studies comprising inelastic collisions of Cl+H$_2$, the barrierless (and
presumably statistical) D$^+$+H$_2$ reaction, and the direct H+D$_2$ reaction
for different final states. Our results show that classical and quantum
deflection functions are essentially coincident whenever quantum interferences
are not preeminent, although the latter are capable of adding valuables
details. When quantum phenomena are present, the quantum deflection function
arises as a powerful tool and makes possible to observe the interference
pattern at first sight, allowing us to disentangle the partial waves that
contribute to constructive and destructive interferences. It also provides
information on the number and position of the peaks in the DCS, something that
cannot be extracted from the classical deflection function. The methodology devised here is
completely general, and can be used to obtain deflection functions for
polyatomic systems. Moreover, it must be stressed that due to its quantum
mechanical nature, it could be used to analyse reaction mechanisms that do not
have a classical analogue or under conditions where the classical deflection cannot be
calculated, such as at energies below the barrier or whenever either resonances
or Fraunhofer diffraction are observed.


\section{Acknowledgment}

The authors acknowledge funding by the Spanish Ministry of Science and
Innovation (grant MINECO/FEDER-CTQ2015-65033-P).  PGJ acknowledges the Spanish
Ministry of Economy  and Competitiveness for the Juan de la Cierva fellowship
(IJCI-2014-20615).

\providecommand*{\mcitethebibliography}{\thebibliography}
\csname @ifundefined\endcsname{endmcitethebibliography}
{\let\endmcitethebibliography\endthebibliography}{}

\end{document}